\documentclass[conference]{IEEEtran}
\IEEEoverridecommandlockouts
\usepackage{cite}
\usepackage{amsmath,amssymb,amsfonts}
\usepackage{algorithm}
\usepackage{algorithmic}
\usepackage{graphicx}
\usepackage{subfigure}
\usepackage{textcomp}
\usepackage{xcolor}
\usepackage{amsthm}
\usepackage{mathrsfs}
\usepackage{bm}
\usepackage{ragged2e}
\usepackage{stfloats}

\def\BibTeX{{\rm B\kern-.05em{\sc i\kern-.025em b}\kern-.08em
    T\kern-.1667em\lower.7ex\hbox{E}\kern-.125emX}}
\begin{document}

\title{Client Selection and Bandwidth Allocation for Federated Learning: An Online Optimization Perspective}


\author{\IEEEauthorblockN{Yun Ji\textsuperscript{1},
Zhoubin Kou\textsuperscript{1},
Xiaoxiong Zhong\textsuperscript{1,2}, 
Hangfan Li\textsuperscript{1},
Fan Yang\textsuperscript{1}, and
Sheng Zhang\textsuperscript{1,*}}
\IEEEauthorblockA{\textsuperscript{1}Graduate School in Shenzhen, Tsinghua University, Shenzhen, 518055, China}
\IEEEauthorblockA{\textsuperscript{2}Peng Cheng Laboratory, Shenzhen 518000, P.R. China}
\IEEEauthorblockA{\textsuperscript{*}Corresponding author: Sheng Zhang, email: zhangsh@sz.tsinghua.edu.cn}}

\maketitle

\begin{abstract}
Federated learning (FL) can train a global model from clients’ local data set, which can make full use of the computing resources of clients and performs more extensive and efficient machine learning on clients with protecting user information requirements. Many existing works have focused on optimizing FL accuracy within the resource constrained in each individual round, however there are few works comprehensively consider the optimization for latency, accuracy and energy consumption over all rounds in wireless federated learning. Inspired by this, in this paper, we investigate FL in wireless network where client selection and bandwidth allocation are two crucial factors which significantly affect the latency, accuracy and energy consumption of clients. We formulate the optimization problem as a mixed-integer problem, which is to minimize the cost of time and accuracy within the long-term energy constrained over all rounds. To address this optimization, we propose the Per-round Energy Drift Plus Cost (PEDPC) algorithm in an online perspective, and the performance of the PEDPC algorithm is verified in simulation results in terms of latency, accuracy and energy consumption in IID and NON-IID dat distributions. 
\end{abstract}


\section{Introduction}
With the exponential growth of smart devices, a huge amount of data has generated each day, which inspires the application of machine learning(ML), such as predicting the traffic for autonomous vehicles, analyzing the health condition of patients [1] and detecting the status of smart home [2]. Federated learning(FL) is one of the most popular directions of distribute ML. Unlike the traditional ML, FL doesn’t centralize distributed data in clients, rather the learning process is distributed in clients using their own data, which protects the user privacy to a large extent [3]. Specifically, FL process can be described below:1): At the beginning of one round, a set of clients are selected by cloud server, then the server distributes global model to these clients; 2): After receiving the global model, the clients perform local training on their own local dataset and transmit the updated model to the cloud server; 3) The cloud server aggregate the transmitted models and update global model.

In wireless network, optimizing FL faces many challenges. On the one hand, the heterogeneous of FL, the computing power of clients varies, which incurs that some stragglers will tremendously prolong the latency of each round. On the other hand, the resource limit, the total bandwidth of all clients are finite in OFDM, which incurs that some clients in a bad channel condition maybe lost data packages while communicating with the server and even become a straggler. Besides, the energy limit, each client has an energy constraint for learning and transmission over all rounds, which incurs that not every client can be selected in each round, and the selection scheme of current round will affect the future selection. Thus optimizing FL is a long-term problem, and it’s difficult to solve. Many existing works mainly consider optimizing the FL accuracy within a resource limit in each individual round, however the latency and energy consumption are as important as the accuracy in a real-world federated learning, so how to achieve the appropriate tradeoff between these targets is quite important. On the other hand, many existing works only consider the optimal problem in each individual round omitting the dependence of different rounds [4-6], Particularly, early effort on long-term FL optimization is presented in [7]. The author maximizes the linear function which is empirically proportional to the accuracy within the energy limit, but it does not consider the time latency of FL.

In this paper, we jointly optimize the client selection and bandwidth allocation to minimize the latency plus negative accuracy (denoted as cost function in part \uppercase\expandafter{\romannumeral2} function within the long-term energy limit. The main contributions are summarized as follows:
\begin{itemize}
\item Due to the complexity of the long-term problem, we propose a Lyapunov-based algorithm called PEDPC to translate the origin offline problem to a per-round problem. We prove that the PEDPC algorithm can achieve a   tradeoff between cost and energy consumption.
\item To solve the mixed-integer problem we optimize client selection and bandwidth allocation iteratively, and propose an algorithm called Increasing Time-Maximum Client Selection(ITMCS) to select clients according to the ascend order of predicted latency of all clients. 
\item We consider the high heterogeneity of client ability, data distribution and data size in the experiments, and the performance of our algorithms is verified by extensive simulations.
\end{itemize}

\section{System Model and Problem Fomulation}
In this paper, we consider a federated learning system with one cloud server and clients which indexed by the set  . Assume that each client $k\in \mathcal{K}$ has a local dataset   with size ${{D}_{k}}$. We assume that there are$R$rounds totally. In each round $r\le R$, we use ${{x}_{k}}(r)$ to denote weather client $k$ is selected or not: if client $k$ is selected in round $r$ then ${{x}_{k}}(r)=1$, vice versa. Let $x(r)=({{x}_{1}}(r),\cdots,{{x}_{K}}(r))$ collects the overall client selection decisions. Similarly, Assume that we have bandwidth $B$ to be allocated in all. Let ${{b}_{k}}(r)$ represent the allocated bandwidth ratio for client $k$, hence its allocated bandwidth is ${{b}_{k}}(r)B$, and $b(r)=({{b}_{1}}(r),
\cdots,{{b}_{K}}(r))$ collects overall allocated bandwidth ratio. In real-world wireless network, the bandwidth allocated for clients usually can not be arbitrarily small due to a finite resource block size [7], so we define ${{b}_{k}}(r)\ge {{b}_{\min }}$.

\begin{figure}[h]
\centering
\includegraphics[width=0.45\textwidth]{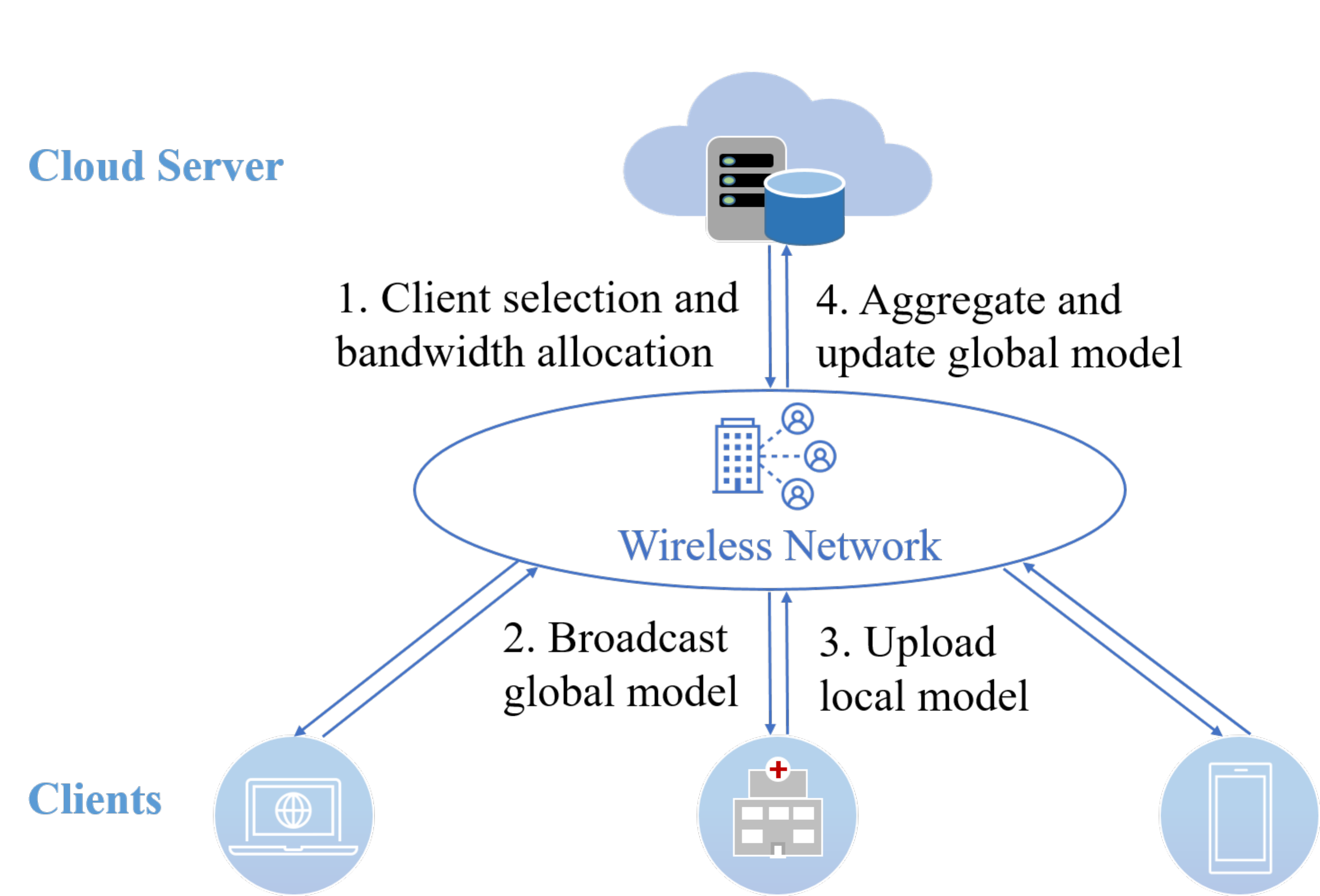}
\caption{Federated learning system model in wireless network}
\label{fig}
\end{figure}
\subsection{Energy Consumption Model}
In a round $r\le R$, all selected clients train local models using their own dataset which incurs the consumption of computational energy. Let ${{f}_{k}}$ and ${{c}_{k}}$denote CPU frequency per second and the number of CPU cycles for training one bit data of client $k$ respectively. Assume that client $k$ conduct ${{U}_{k}}(r)$ training iterations in round $r$. Then the computational energy consumption of client $k$ can be given by :
\begin{equation}
    E_{k}^{\text{cmp}}={{U}_{k}}(r){{\delta }_{k}}{{c}_{k}}{{D}_{k}}(r)f_{k}^{2}\tag{1}
\end{equation}
where ${{\delta }_{k}}$ is the effective capacitance coefficient of computing chipset for client $k$ [4].
After training, the selected clients upload their model parameters to the server. According to the Shannon equation, the transmission rate is:
\begin{equation}
    {{R}_{k}}(r)={{b}_{k}}(r)B\log \left( 1+\frac{{{p}_{k}}(r)h_{k}^{2}(r)}{{{N}_{0}}} \right)\tag{2}
\end{equation}
where ${{p}_{k}}(r)$ is the average transmission power of client $k$ in round $r$ and ${{N}_{0}}$ is the variance of the white Guassian noise. $h_{k}(r)$ denotes the channel gain between client $k$ and the server in round $r$. Let ${{S}_{k}}$ denotes the data size of model parameters, and the communication latency $T_{k}^{\text{com}}(r)$ can be given by:
\begin{equation}
   T_{k}^{\text{com}}(r)=\frac{{{S}_{k}}}{{{R}_{k}}(r)}\tag{3}
\end{equation}

By multiplying transmission power ${{p}_{k}}(r)$ to (4), the communication energy is given by:
\begin{equation}
    E_{k}^{\text{com}}(r)=\frac{{{p}_{k}}(r){{S}_{k}}}{{{R}_{k}}(r)}\tag{4}
\end{equation}

The total energy consumption can be given by:
\begin{equation}
    \begin{aligned}
    {{E}_{k}}(r)
    &=E_{k}^{\text{cmp}}(r)+E_{k}^{\text{com}}(r)\\
    &={{U}_{k}}(r){{\delta }_{k}}{{c}_{k}}{{D}_{k}}(r){{f}_{k}}^{2}+\frac{{{p}_{k}}(r){{S}_{k}}}{{{R}_{k}}(r)}
    \end{aligned}
    \tag{5}
\end{equation}

\subsection{Latency Model}
Let $T_{k}^{\text{cmp}}(r)$ denote the latency of client $k$ on training local model, which is given by:
\begin{equation}
    T_{k}^{\text{cmp}}(r)={{U}_{k}}(r)\frac{{{c}_{k}}{{D}_{k}}}{{{f}_{k}}}\tag{6}
\end{equation}

The total latency of client $k$ in round $r$ can be given by:
\begin{equation}
\begin{aligned}
    {{T}_{k}}(r)
    &=T_{k}^{\text{com}}(r)+T_{k}^{\text{cmp}}(r)\\
    &=\frac{{{S}_{k}}}{{{R}_{k}}(r)}+{{U}_{k}}(r)\frac{{{c}_{k}}{{D}_{k}}(r)}{{{f}_{k}}}
\end{aligned}
\tag{7}
\end{equation}

In FL process, the latency of one round depends on the maximum latency of all selected clients in this round. We denote ${{T}_{0}}(x(r),b(r))$ to represent the latency of round $r$, which can be given by:
\begin{equation}
    {{T}_{0}}(x(r),b(r))=\underset{k=1,\cdots,K}{\mathop{\max }}\,\{{{x}_{k}}(r){{T}_{k}}(r)\}\tag{8}
\end{equation}

\subsection{Accuracy and Cost Model}
According to [4-5,8-9], FL accuracy is largely affected by data size, which can be given by:
\begin{equation}
    \begin{aligned}
    \Phi \left( x(r) \right)
    &=\sum\limits_{k=1}^{K}{\log \left( 1+\mu {{D}_{k}}(r){{x}_{k}}(r) \right)}\\
    &=\sum\limits_{k=1}^{K}{\log \left( 1+{{v}_{k}}{{x}_{k}}(r) \right)}
    \end{aligned}
    \tag{9}
\end{equation}
where $\mu $ is the system parameter. For simplicity, let ${{v}_{k}}=\mu {{D}_{k}}(r)$. In this work, we denote cost function as follows:
\begin{equation}
    {{y}_{0}}(x(r),b(r))={{T}_{0}}(x(r),b(r))-\Phi (x(r))\tag{10}
\end{equation}

\subsection{Problem Formulation}
A joint problem of client selection and bandwidth allocation is considered in this section, which minimizes the cost function over rounds while satisfying energy budget constraints of each client. The problem is formulated as follows:
\begin{align*}
&\textbf {P1}:~\min_{x(0),b(0),...,x(R-1),b(R-1)} \frac{1}{R}\sum\limits_{r=0}^{R-1}{{{y}_{0}}(x(r),b(r))}\tag{11}\\
&{\qquad\text { s.t.}}~\sum\limits_{r=0}^{R-1}{{{x}_{k}}(r){{E}_{k}}(r)}\le {{H}_{k}},\forall k \tag{12}\\
&\hphantom {{\qquad\text { s.t.}}~}{{b}_{k}}(r)\ge {{b}_{\min }},\forall k,\forall r \tag{13}\\
&\hphantom {{\qquad\text { s.t.}}~}\sum\limits_{k=1}^{K}{{{b}_{k}}(r)}=1,\forall r \tag{14}\\
&\hphantom {{\qquad\text { s.t.}}~}{{x}_{k}}(r)\in \{1,0\},\forall k,\forall r\tag{15}
\end{align*}

where constraint (12) guarantees that the energy consumption of selected client $k$ over $R$ rounds does not exceed the energy budget ${{H}_{k}}$. Constraint (13) ensures that the bandwidth of every client is at least ${{b}_{\min }}$. Constraint (14) indicates that the sum of all allocated bandwidth equals to $B$. Constraint (15) defines that each client is selected or not in round $r$. 

To solve \textbf{P1}, we have to know full information (e.g., channel gain) about the future $R$ rounds in advance, but it’s almost impossible in practical situation. Thus, we need to translate P1 into an online problem which can be solved without knowing any prior knowledge of future rounds. First we divided $R$ rounds into $F$ frames, and each frame includes $L=R/F$ rounds, then we translate \textbf{P1} to a series of \textbf{P2} for $f=1,2,...,F$.
\begin{align*}
&\textbf{P2}:~\min_{x(r),b(r)} {{c}_{f}}\triangleq \frac{1}{L}\sum\limits_{r=fL}^{(f+1)L-1}{{{y}_{0}}(x(r),b(r))}\tag{16}\\
&\qquad {\quad\text { s.t.}}~\sum\limits_{r=fL}^{(f+1)L-1}{{{x}_{k}}(r){{E}_{k}}(r)}\le {{H}_{k}}/F,\forall k \tag{17} \\
&\qquad\qquad\quad\text{Constraints} (13), (14), (15)
\end{align*}

It can be found that \textbf{P2} is still an offline problem, and we denote $c_{f}^{*}$ as the optimal value of (16) by L-round lookahead algorithm, which finds the optimal solution by foreseeing full offline knowledge over the frame in advance [8]. We assume throughout that the constraints are feasible for \textbf{P2}, such as there is no client is selected in any round. Therefor we use $\frac{1}{F}\sum_{f=0}^{F-1}{c_{f}^{*}}$ to approach the optimal value of P1.

\section{Online Optimization Algorithm and Convergence Analysis}
In this section, we develop the Per-round Energy Drift Plus Cost algorithm, called PEDPC, to turn \textbf{P2} to a series of online problems and develop the iterative algorithm to select clients and allocate bandwidth iteratively. Specifically we propose the Increasing Time-Maximum Client Selection algorithm to select clients, called ITMCS, and adopt the Barrier Method algorithm to allocate bandwidth for the selected clients at each round.  Fig. 2 demonstrates the problem transform and algorithm framework.

\subsection{The PEDPC Algorithm}
One major challenge of solving \textbf{P2} is that constraint (17) couples long-term energy restrict which need to be decoupled into per-round term. For this, we apply the Lyapunov theory [11] and introduce a virtual queue ${{Z}_{k}}=\{{{Z}_{k}}(0),{{Z}_{k}}(1),\cdots,{{Z}_{k}}(R-1)\}$ for client $k\in \mathcal{K}$. Besides, we denote the queue backlog vector as $Z(r)=\{{{Z}_{1}}(r),{{Z}_{2}}(r),\cdots,{{Z}_{K}}(r)\}$ in round $r$. Each queue is configured as follows:
\begin{equation}
    {{Z}_{k}}(r+1)=\max[{{Z}_{k}}(r)+{{x}_{k}}(r){{E}_{k}}(r)-{{H}_{k}}/R,0]\tag{18}
\end{equation}
where ${{Z}_{k}}(r)$ is the queue backlogs of client $k$ in round $r$, and represents the difference between the energy consumption and budget of client $k$over $r$ rounds. According to [11], the Lyapunov function $Y(Z(r))$ can be given by:
\begin{equation}
    Y(Z(r))=\frac{1}{2}\sum\limits_{k=1}^{K}{{{Z}_{k}}{{(r)}^{2}}}\tag{19}
\end{equation}

This represents a scalar measure of queue congestion in the network. Besides, we denote the energy drift function as the change of Lyapunov function between two rounds: $Y(Z(r+1))-Y(Z(r))$, and an upper bound can be given by \textbf{Lemma 1}.

\textbf{Lemma 1}: Assume $y_{k}^{\min }$, $y_{k}^{\max }$ are two constants, and for $\forall k\in \mathcal{K}$, we have $y_{k}^{\min }\le {{x}_{k}}(r){{E}_{k}}(r)-{{H}_{k}}/(LF)\le y_{k}^{\max }$, then:
\begin{equation}
    \begin{aligned}
    & Y(Z(r+1))-Y(Z(r)) \\ 
    \le & D+\sum\limits_{k=1}^{K}{{{Z}_{k}}(r)}({{x}_{k}}(r){{E}_{k}}(r)-{{H}_{k}}/(LF)) \\ 
    \end{aligned}
    \tag{20}
\end{equation}
where $D=\frac{1}{2}\sum\limits_{k=1}^{K}{\max [{{(y_{k}^{\min })}^{2}},{{(y_{k}^{\max })}^{2}}]}$.

\textbf{Proof} : see appendix A.

Now we formulate \textbf{P3}: in order to decoupled the constraint (17), we construct the objective of \textbf{P3} by adding the upper bound of energy drift function to the per-round objective of \textbf{P2} since the fact that $Y(Z(r+1))-Y(Z(r))$ being “small” for each round $r\le R$ implies the energy deficit of all clients being “small”. Thus \textbf{P3} can be given by:
\begin{align*}
&\textbf{P3}:~\min_{x(r),b(r)} \sum\limits_{k=1}^{K}{{{Z}_{k}}(r)}({{x}_{k}}(r){{E}_{k}}(r)-{{H}_{k}}/R)\\
&\qquad \qquad \qquad +V{{y}_{0}}(x(r),b(r))+D\tag{21}\\
&\qquad {\quad\text { s.t.}}~\text{Constraints} (13), (14), (15) 
\end{align*}
where $D$ is a constant which can be omitted in optimizing. $V>0$ is a control parameter: when $V$ is close to zero, reducing energy consumption is more important, otherwise, reducing the cost function ${{y}_{0}}(x(r),b(r))$ is more crucial.

 According to the above, we develop the PEDPC algorithm to solve \textbf{P2} in an online way.

\renewcommand{\algorithmicrequire}{ \textbf{Input:}} 
\renewcommand{\algorithmicensure}{ \textbf{Output:}} 
\begin{algorithm}
\caption{ PEDPC}
\label{alg:Framwork}
\begin{algorithmic}[1] 
\REQUIRE ~~\\ 
    $Z_{k}(0)=0, \forall k$, $L$ and $F$;\\
    \FOR{$f=0,1, \cdots, F-1$}
        \FOR{$r=0,1, \cdots, L-1$}
            \STATE Observe the current energy deficit $Z_{k}(r),  \forall k$ and the channel state $h_{k}(r), \forall k$  
            \STATE Solve P3;\
            \STATE Update queue backlog vector $Z(r+1)$ according to (18);\
        \ENDFOR  
        \STATE $V \leftarrow V_{f}$
    \ENDFOR
\end{algorithmic}
\end{algorithm}
 
 PEDPC algorithm decouple \textbf{P2} to a series of per-round problem, in each round, only requires the current energy deficit and channel state to finish selecting clients and allocating bandwidth without foreseeing any future information. After each frame, we update the value of   to dynamically adjust the key factors of \textbf{P3}, and more detailed impact of   will be discussed in simulation results. 
 
 \subsection{The Iterative Algorithm}
By considering the second term $\sum_{k=1}^{K}{{{Z}_{k}}(r)}({{x}_{k}}(r){{E}_{k}}(r)-{{H}_{k}}/(R))$of (21), we can found that if one client is selected often, its energy queue backlog ${{Z}_{k}}$ will be large which incurs that to reduce the energy deficit is more significant. Besides, if one client’s energy consumption exceeds the budget, then select this client less to decrease energy consumption. Plugging (5), (7), (8), (9) into (21) yields:
\begin{align*}
&\mathbf {P3.eq}:~\min_{x(r),b(r)} \left[ {{x}_{k}}(r)\left( T_{k}^{\text{cmp}}(r)+\frac{{{S}_{k}}}{{{b}_{k}}(r){{G}_{k}}(r)} \right) \right]\\
&\qquad \qquad \qquad -V\sum\limits_{k=1}^{K}{\log \left( 1+{{v}_{k}}{{x}_{k}}(r) \right)}\\
&\qquad \qquad \qquad +V{{y}_{0}}(x(r),b(r))+D\tag{22}\\
&\qquad {\quad\text { s.t.}}~\text{Constraints} (13), (14), (15) 
\end{align*}
where Notice that \textbf{P3.eq} is a mixed-integer problem and there is almost no algorithm to get the optimal solution in polynomial time. So we develop Iterative algorithm with low complexity in \textbf{algorithm 2} to solve \textbf{P3.eq}.

\begin{algorithm}
\caption{ Iterative Algorithm}
\label{alg:Framwork}
\begin{algorithmic}[1] 
\REQUIRE ~~\\ 
    A feasible solution ($\bm{ x^0(r) },\bm{b^0(r)}$) of P3.eq ,$i=0$ and $I>0$;\\
    \REPEAT
        \STATE With given $\bm{b^0(r)}$, obtain the optimal $\bm{ x^{i+1}(r) }$ of P4;
        \STATE With given $\bm{ x^{i+1}(r) }$, obtain the optimal $\bm{b^{i+1}(r)}$ of P5;
        \STATE set $i=i+1$;
    \UNTIL $i\geq{I}$
\end{algorithmic}
\end{algorithm}

Iterative algorithm contains two steps at each iteration: To solve \textbf{P3.eq}, we first fix one variable $b(r)$ and optimize $x(r)$ which is a 0-1 integer problem given in P4. Then we optimize $b(r)$ with fixed updated $x(r)$, which is a convex problem given in \textbf{P5}, The advantage of Iterative algorithm is that the objective value of \textbf{P3.eq} is non-increasing in each step and it always converges to a local optimal solution since the objective value is lower bounded by zero when we choose no client in round $r$.

\subsection{The ITMCS Algorithm}
The client selection problem \textbf{P4} can be given by:
\begin{align*}
&\textbf{P4}:~\min_{x(r)} V \max_{k\in \mathcal{K}}\left[ {{T}_{k}}(r){{x}_{k}}(r) \right]\tag{23}\\
&\qquad \qquad +\sum\limits_{k=1}^{K}{\left( Z_{k}^{'}(r){{x}_{k}}(r)-V\log \left( 1+{{v}_{k}}{{x}_{k}}(r) \right) \right)}\\
&\qquad {\quad\text { s.t.}}~\text{Constraints} (15)
\end{align*}
where ${{Z}_{k}}(r)\left( E_{k}^{cmp}(r)+\frac{{{p}_{k}}(r){{S}_{k}}}{{{b}_{k}}(r){{G}_{k}}(r)} \right)=Z_{k}^{'}(r)$. We propose the Increasing Time-Maximum Client Selection, called ITMCS, to efficiently figure out the optimal solution of \textbf{P4}.

In ITMCS, we let ${{q}_{k}}(r)=Z_{k}^{'}(r){{x}_{k}}(r)-V\log \left( 1+{{v}_{k}}{{x}_{k}}(r) \right)$, and denote a set ${{S}_{0}}$ that collects all clients which satisfy ${{q}_{k}}(r)<0$. Then we rank the clients in ${{S}_{0}}$ according to their time latency, and add clients one by one in an ascending order. Until we go through the entire ${{S}_{0}}$ and obtain the selection set of ${{S}^{i}}$ for $i=1,2,\cdots,|{{S}^{0}}|$, where $|{{S}^{0}}|$ means the size of ${{S}_{0}}$. we denote $\mathcal{S}=\{{{S}^{i}}\},i=1,2,\cdots,|{{S}^{0}}|$, and we figure out the optimal clients selected set ${{S}^{*}}=\arg \min_{s\in \mathcal{s}}\,V{{T}_{k}}(r)+\sum_{k\in S}{{{q}_{k}}}$.We assume there are $m$ clients being selected, i.e., $|{{S}^{*}}|=m$.

\begin{algorithm}
\caption{Increasing Time-Maximum Client Selection}
\label{alg:2}
\begin{algorithmic}[1]
\REQUIRE ~~\\ 
    $x_{k}\left(r\right) = 0, \forall{k}$;\\
    \STATE Set $S_{0}=\varnothing, S=S_{0}, \mathcal{S}=\left\{S_{0}\right\}$;\
    \STATE Calculate $q_{k}=Z_{k}^{\prime}(r)-V \log \left(1+v_{k}\right), \forall k$;\
    \STATE Find $k$ to satisfy $q_{k}<0, \forall k$ and Update $S_{0}=S_{0} \cup\{k\}$;\
    \STATE Rank the clients in $S_{0}$ according to $T_{k}(r)$. Hence we have $T_{1}(r) \leq T_{2}(r) \leq \cdots \leq T_{\left|S_{0}\right|}(r)$;\
    \FOR{$i \in S_{0}$}
        \STATE Update $S=S \cup\{k\}$, where $T_{k}(r) \leq T_{i}(r), \forall k \in S_{0}$;\
        \STATE Update $\mathcal{S}=\mathcal{S} \cup\{S\}$;\
        \STATE Calculate $W(S)= V T_{k}(r)+\sum_{k \in S} q_{k}$;\
        \STATE Set $S=\varnothing$;\
    \ENDFOR
    \STATE Find $S^{*}=\arg \min _{S \in \mathcal{S}}(W(S))$;\
    \STATE Return $\boldsymbol x^{*}$, where $x_{k}^{*}=1\left\{k \in S^{*}\right\}, \forall k$;
\end{algorithmic}
\end{algorithm}

\subsection{The Barrier Method Algorithm for Bandwidth Allocation}
The bandwidth allocation problem \textbf{P5} can be given by:
\begin{align*}
&\textbf{P5}:~\min_{x(r)} V \max_{k\in {{S}^{*}}}\left[ T_{k}^{\text{cmp}}(r)+\frac{{{S}_{k}}}{{{b}_{k}}(r){{G}_{k}}(r)} \right]\\
&\qquad \qquad \qquad +\sum\limits_{k\in {{S}^{*}}}^{{}}{\frac{{{p}_{k}}(r){{Z}_{k}}(r){{S}_{k}}}{{{b}_{k}}(r){{G}_{k}}(r)}}\tag{24}\\
&\qquad {\quad\text { s.t.}}~\text{Constraints} (13),(14)
\end{align*}

It is difficult to handle the max-term of (24) because it’s non-differentiable, and we replaced it with $\ln \left[ \sum_{k\in {{S}^{*}}}{\exp \left( T_{k}^{\text{cmp}}(r)+\frac{{{S}_{k}}}{{{b}_{k}}(r){{G}_{k}}(r)} \right)} \right]$ due to the inequality $\max \{ {{x}_{1}},{{x}_{2}},...,{{x}_{m}} \}\le \ln ({{e}^{{{x}_{1}}}}+{{e}^{{{x}_{2}}}}+...+{{e}^{{{x}_{m}}}})\le \max ({{x}_{1}},{{x}_{2}},...,{{x}_{m}})+\ln (m)$ , which indicates the absolute error $e\le \ln (|{{S}_{r}}|)$. Thus \textbf{P6} is given by:
\begin{align*}
&\textbf{P6}:~\min_{b(r)} V \ln \left[ \sum\limits_{k\in {{S}^{*}}}{\exp \left( T_{k}^{\text{cmp}}(r)+\frac{S_{k}^{'}(r)}{{{b}_{k}}(r)} \right)} \right]\\
&\qquad \qquad \qquad +\sum\limits_{k\in {{S}^{*}}}^{{}}{\frac{G_{k}^{'}(r)}{{{b}_{k}}(r)}}\tag{25}\\
&\qquad {\quad\text { s.t.}}~\text{Constraints} (13),(14)
\end{align*}
where $\frac{{{S}_{k}}}{{{G}_{k}}(r)}=S_{k}^{'}(r),\frac{{{p}_{k}}(r){{Z}_{k}}(r){{S}_{k}}}{{{G}_{k}}(r)}=G_{k}^{'}(r)$

\textbf{Theorem 1}: \textbf{P6} is a convex problem.

\textbf{Proof} : see appendix B.

We adopt the well-known Barrier method to solve \textbf{P6}, more details about Barrier method are given by chapter 11 of [12].

\section{Algorithm Convergence and Complexity Analysis}
In this section, we analyze the convergence and time complexity of our proposed algorithms.

\subsection{Convergence Analysis}
\textbf{Assumption 1}: For $\forall k$, $y_{k}^{\min }\le {{x}_{k}}(r){{E}_{k}}(r)-{{H}_{k}}/R\le y_{k}^{\max }$ where $y_{k}^{\min }$ and $y_{k}^{\max }$ are constants and $L>0$, $F>0$. Assume the initial queue backlog vector $Z(0)$ is finite.

\textbf{Definition 1}: A discrete time queue $Q(t)$ is mean rate stable if:
\begin{equation}
    \underset{r\to \infty }{\mathop{\lim }}\,\frac{\mathbb{E}\{|Q(t)|\}}{t}=0\tag{26}
\end{equation}

\textbf{Lemma 2}: For any client $k\in \mathcal{K}$, we have:
\begin{equation}
   {{Z}_{k}}(R)-{{Z}_{k}}(0)\ge \sum\limits_{r=0}^{R-1}{{{x}_{k}}(r){{E}_{k}}(r)-{{H}_{k}}}  \tag{27}
\end{equation}
                       
\textbf{Proof} : see appendix C.

\textbf{Theorem 2}: If the PEDPC algorithm is implemented every round, compared with L-round lookahead algorithm, the following inequality holds:
\begin{equation}
  \frac{1}{R}\sum\limits_{r=0}^{FL-1}{{{y}_{0}}(x(r),b(r))}\le \frac{1}{F}\sum\limits_{f=0}^{F-1}{c_{f}^{*}}+\frac{DL}{V}+\frac{Y(Z(0))}{VR}\tag{28}
\end{equation}

\textbf{Proof} : see appendix D.

\textbf{Theorem 2} implies that the difference between objective value by PEDPC algorithm and the optimal value is upper bounded by $\frac{DL}{V}+
\frac{Y(Z(0))}{VR}$. If the initial queue backlog vector $Z(0)=0$, then the final term $\frac{Y(Z(0))}{VR}$ equals 0. Thus we have that the objective of P1 is within $O(1/V)$ of the optimal value. Now we compare the energy consumption of drift-plus-penalty algorithm with the L-round lookahead algorithm, and the following theorem holds:

\textbf{Theorem 3}: If the PEDPC algorithm is implemented every round, the virtual queue $Z(r)$ is mean rate stable and the total energy consumption of client $k$ is upper bounded by${{H}_{k}}+\sqrt{2DRL+2VL\sum_{f=0}^{F-1}{(c_{f}^{*}}-y_{0}^{\min })}$.

\textbf{Proof} : see appendix E.

According to \textbf{Theorem 2} and \textbf{Theorem 3}, we can found that the PEDPC algorithm can achieve $[O(1/V),O(\sqrt{V})]$ tradeoff between cost and energy consumption in our system model.

\subsection{Complexity Analysis}
To solve the origin problem \textbf{P1} by using \textbf{Algorithm 1}, the major complexity depends on solving P5. According to [12], the time complexity of the barrier method is $O\left( \left\lceil \frac{\log \left( 2m/({{\varepsilon }_{br}}{{\alpha }^{(0)}}) \right)}{\log \mu } \right\rceil  \right)$. The complexity of \textbf{Algorithm 2} is $O(m)$, hence the time complexity of solving our proposed problem is $O\left( RIm\left\lceil \frac{\log \left( 2m/({{\varepsilon }_{br}}{{\alpha }^{(0)}}) \right)}{\log \mu } \right\rceil  \right)$. We can found that the complexity grows linearly with the product of the number of clients and FL iterative rounds.

\section{Simulation Results}

\subsection{Experiment Settings}
In our simulation, we consider the hand-written digit classification task on MNIST dataset [13]. For MNIST dataset, we apply a multi-layer perceptron which has two hidden layers with 10 hidden nodes each, and the number of parameters is 7960 (model parameter size ${{S}_{k}}=0.24Mbits$ in 32-bit float). Besides, we consider two data distribution cases: 1) IID case, where the data of training set is shuffled and uniformly distributed over all clients. 2) NON-IID case, where the data size of each client is uniformly distributed in [1.2, 2.4, 3.6, 4.8, 6] Mbits. In addition, we sort the training data by the labels and divide them into 300 groups, every client choose 1-5 groups according to its own data size and at most contains five digits. FedAvg [3] is used as the learning algorithm in FL. The parameter settings are shown in TABLE I.
\begin{table}[h]
\renewcommand\arraystretch{1.5}
\caption{Simulation Parameters}
\begin{center}
\begin{tabular}{c|c}
\hline
\textbf{Parameters}&    \textbf{Value} \\
\hline
Number of clients, $K$ & $100$\\
\hline
Number of global rounds,$R$ & 300\\
\hline
Number of local training iterative, ${{U}_{k}}(r)$ & 5\\
\hline
CPU cycles  for training 1 bit data, ${{c}_{k}}$ & $1\sim 10$ cycles/bit\\
\hline
CPU frequency, ${{f}_{k}}$ & $0.01\sim 1GHz$\\
\hline
Effective switched capacitance, ${{\delta }_{k}}$& ${{10}^{-28}}$\\
\hline
System bandwidth, $B$ & 10MHz\\
\hline
Square of channel gain, $h_{k}^{2}(r)$ & ${{10}^{-9}}\sim {{10}^{-11}}$\\
\hline
Transmission power, ${{p}_{k}}(r)$ & $10\sim 20$dBm\\
\hline
Noise power, ${{N}_{0}}$ & ${{10}^{-13}}W$\\
\hline
System parameter, $\mu $ & $1.7\times {{10}^{-8}}$\\
\hline
Energy budget, ${{H}_{k}}$ & 1.5J\\
\hline
\end{tabular}
\label{tab1}
\end{center}
\end{table}

Furthermore, to verify the effectiveness of our proposed algorithm we consider the following four benchmark algorithms:

\textbf{Select all}: All clients are selected in each round, and the bandwidth is allocated equally.

\textbf{Select randomly}: For a given probability$Pr$, randomly select $Pr*n$ clients in each round, and the bandwidth is allocated equally for the selected clients, which is also called Randomly algorithm in our experiments.
\begin{figure*}[htb]
\centering
\subfigure[ Average number of clients]{\includegraphics[width=0.3\textwidth]{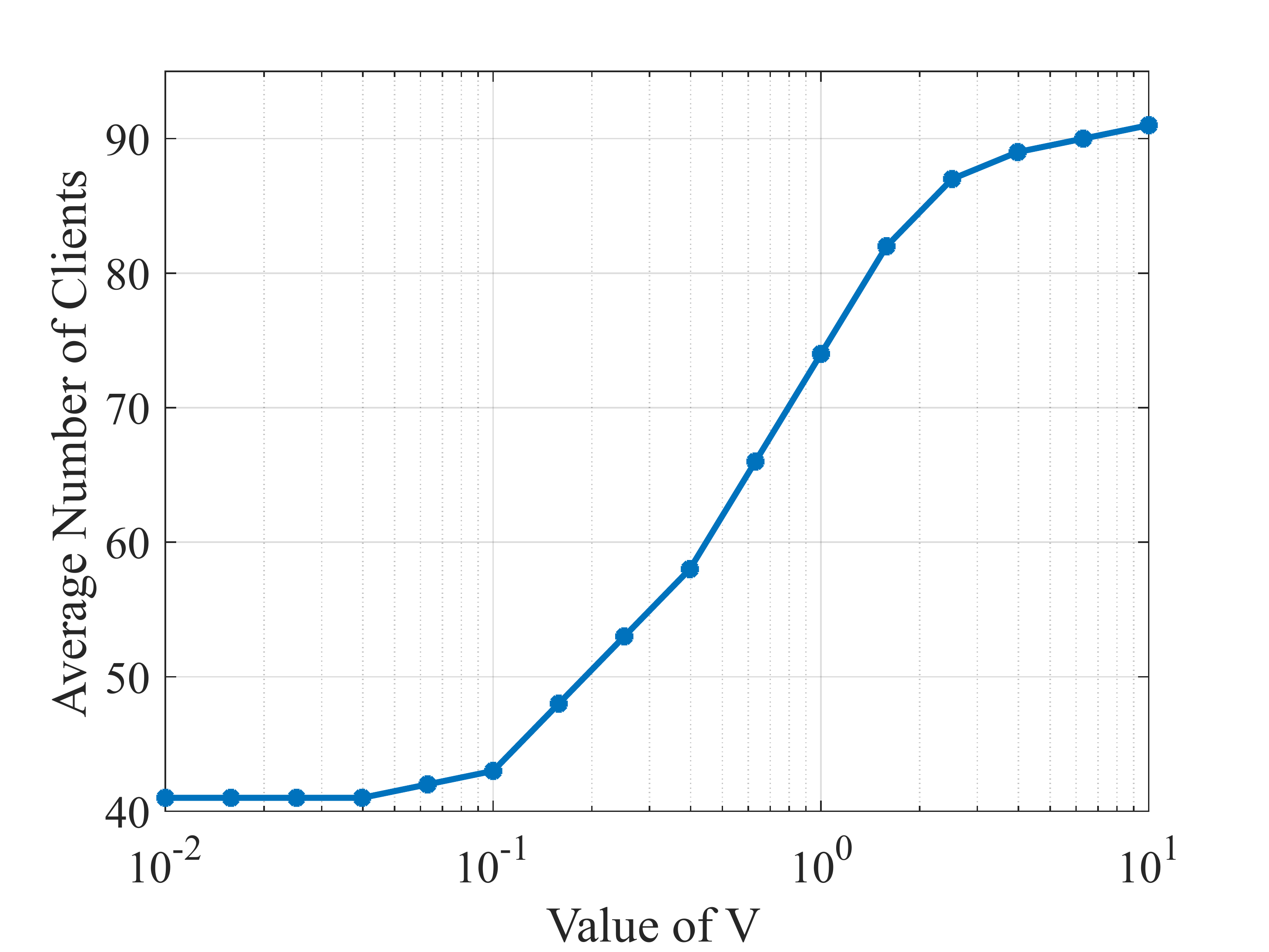}\label{fig: 1}}
\quad
\subfigure[Time latency]{\includegraphics[width=0.3\textwidth]{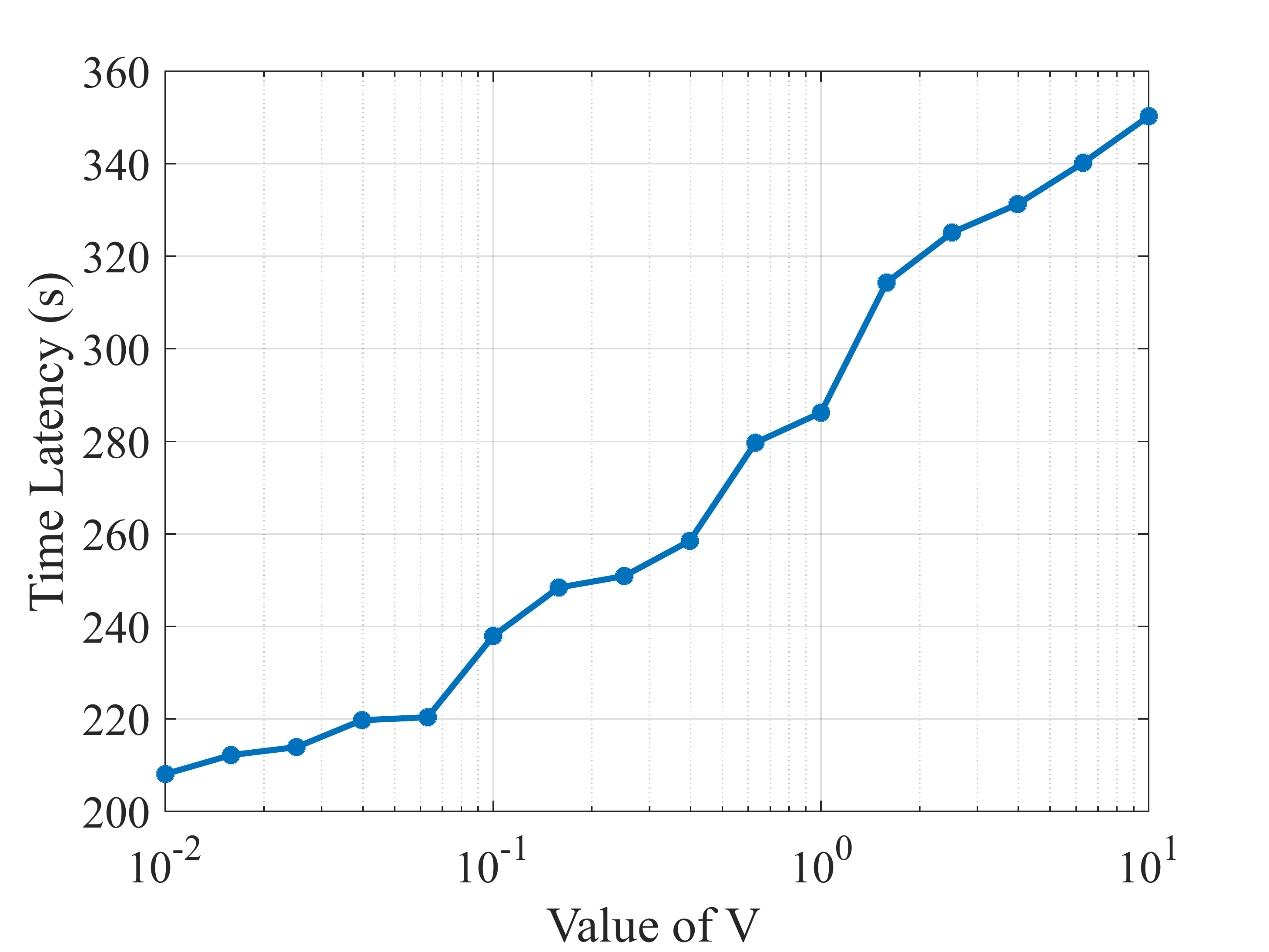}\label{fig: 2}}
\quad
\subfigure[ Energy Overflow]{\includegraphics[width=0.3\textwidth]{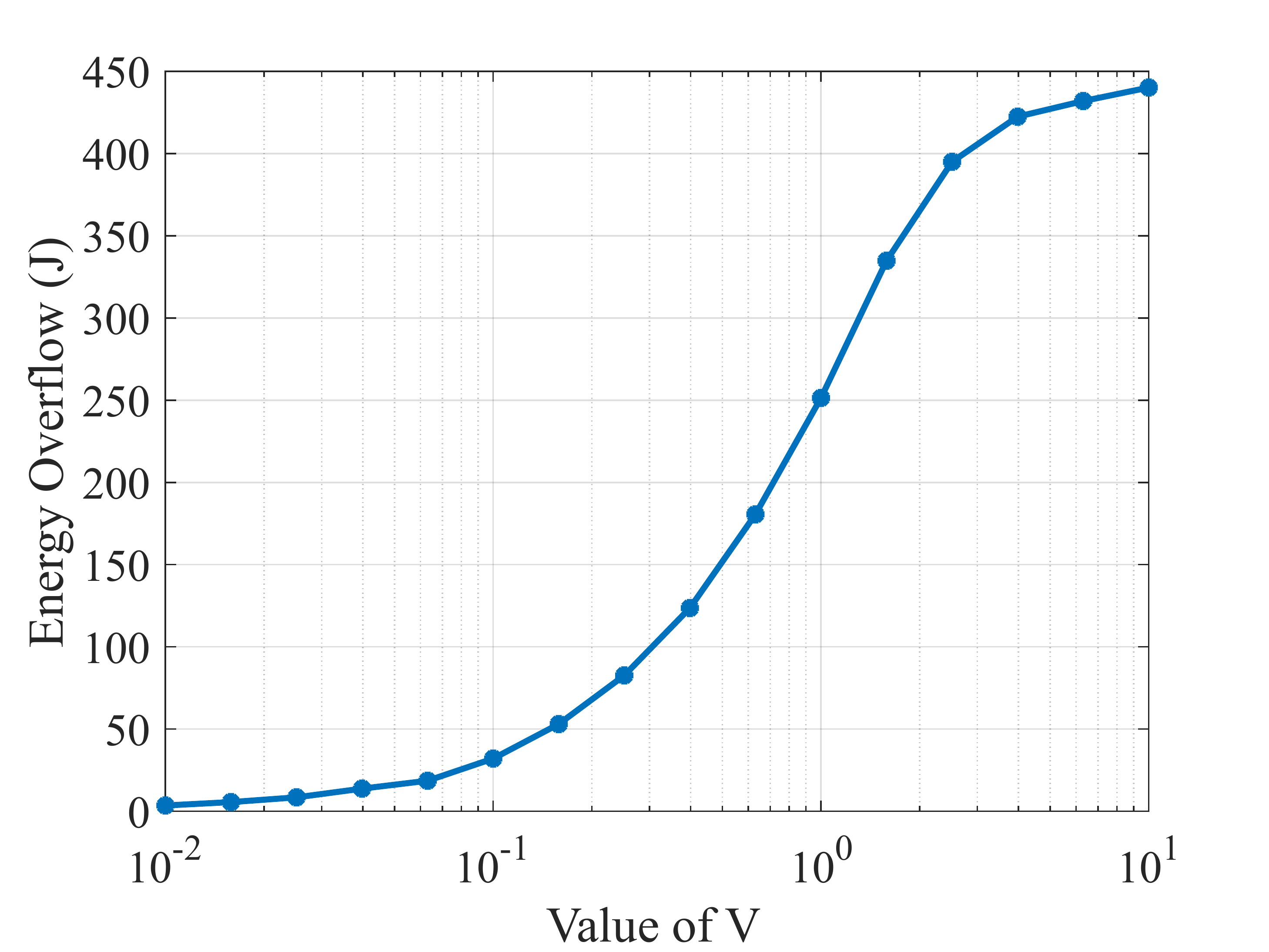}\label{fig: 3}}
\caption{ The impact of V on the average number of clients, time latency and energy overflow}
\label{fig1}
\end{figure*}

\textbf{Select greedily}: Choose as many as clients within an energy limit, which is also called Greedily algorithm in our experiments. In each round, select clients by solving:
\begin{align*}
& \qquad ~\max_{x(r),b(r)}\,\sum\limits_{k=1}^{K}{{{x}_{k}}(r)}\tag{29}\\
&\qquad {\quad\text { s.t.}}~\text{ }{{x}_{k}}(r){{E}_{k}}(r)\le {{H}_{k}}/R,\forall k,\forall r\tag{30}\\
&\qquad {\qquad\quad}\text{Constraints} (13),(14),(15)
\end{align*}

This problem can be solved by:1) according to ${{E}_{k}}(r)={{H}_{k}}/R$ calculate the bandwidth vector of all clients ${{b}^{g}}(r)=\{b_{1}^{g}(r),b_{2}^{g}(r),...,b_{k}^{g}(r)\}$; 2) set ${{S}^{g}}=\varnothing $ and $b_{i}^{g}(r)={{b}_{\min }}$, for $i\in \{i|b_{i}^{g}\le {{b}_{\min }}\}$; 3) rank ${{b}^{g}}(r)$ in the ascending order and add client one by one to ${{S}^{g}}$ until the total bandwidth of all clients of ${{S}^{g}}$exceeds $B$. 4) set ${{x}_{k}}(r)=1$ for $k\in {{S}^{g}}$, and other clients are not selected.

\textbf{FedCS} [14]: Select as many as clients within an time latency limit. In each round, select clients by solving:
\begin{align*}
&\qquad ~\max_{x(r),b(r)}\,\sum\limits_{k=1}^{K}{{{x}_{k}}(r)}\tag{31}\\
&\qquad{\quad\text { s.t.}}\text{ }{{x}_{k}}(r){{T}_{k}}(r)\le {{T}_{\max }},\forall k,\forall r\tag{32}
\end{align*}
the solution is similar as the Greedily algorithm.
\subsection{Impact of $V$}
In \textbf{PEDPC}, $V$ is a vital parameter to control the key optimizing factors of \textbf{P3}. In this section, we simulate the impact of $V$ on the average number of clients, latency and energy overflow in FL. Specially energy overflow equals the energy consumption minus the energy budget).

Fig. 2 shows the trend of average number of  selected clients, time latency and energy overflow as $V$ growing. As we can see, a large $V$ emphasizes more on the cost function resulting in more clients selected and slow growth time latency. Oppositely, a small $V$ emphasizes more on the energy consumption, which incurs a small overflow on the energy budget.

\subsection{Performance Comparison}
To verify the performance of \textbf{PEDPC}, we experiment under IID and NON-IID data distributions.


Fig. 3 shows the energy overflow, time latency and accuracy of \textbf{PEDPC} and benchmarks in IID case. For comparison purpose, we set the average client number of above algorithms to be 40 other than \textbf{Select all}. Compared with \textbf{Greedily}, \textbf{PEDPC} saves about $5.5\sim 6.5$ times energy and latency, meanwhile achieves a higher accuracy. For selecting more clients, \textbf{Greedily} assigns each client the minimum bandwidth meeting the energy constraint which leads to a huge consumption of latency. Compared with \textbf{Randomly}, we can found that \textbf{PEDPC} saves about $5.8\sim8.0$ times energy and time latency while achieving a higher accuracy. This is because \textbf{Randomly} distributes bandwidth evenly for all selected clients, and takes no account of the consumption of energy and time, which incurs the energy and latency a rise. Even though \textbf{FedCS} saves three times latency of our proposed algorithm, it uses 26.6 times energy of ours. This is because \textbf{FedCS} takes no account of energy consumption. Because Select all algorithm picks up all clients, its energy consumption and time latency are much larger than ours.




Fig. 4 shows the energy overflow, time latency and accuracy of \textbf{PEDPC} and benchmarks in NON-IID case. For comparison purpose, we set the average number of clients 90 other than \textbf{Select all}. Because of the similar number of selected clients, the accuracy of all algorithms does not differ by more than 3 percents. Compared with \textbf{Greedily} and \textbf{Randomly}, the time latency of \textbf{PEDPC} achieves $33\%$  of above two algorithms under a similar energy consumption. This is because the bandwidth allocation of these two algorithms without regarding to the latency. As the number of clients increasing, the advantage of \textbf{FedCS} in saving time gradually diminishes, and the energy consumption is 4.5 times of ours since the bandwidth allocation policy omits the energy consumption. \textbf{Select all} takes much more energy and time latency than ours.

\section{Conclusion}
In this paper, we propose a joint optimization algorithm PEDPC with considering client selection and bandwidth allocation for wireless federated leaning. In PEDPC, We exploit the Lyapunov-based energy deficit queue to solve the optimization problem with ITMCS algorithm from an online optimization perspective. Then we prove that PEDPC can achieve an $[O(1/V), (\sqrt{V})]$ cost-energy tradeoff. The extensive simulations show PEDPC achieves a better performance comparing with {Sellect all},  {Randomly}, {Greedily}, and {FedCS} in terms of energy consumption, time latency and FL accuracy.

\begin{figure*}[htb]
\footskip = 30pt
\centering
\subfigure[Energy overflow]{\includegraphics[width=0.3\textwidth]{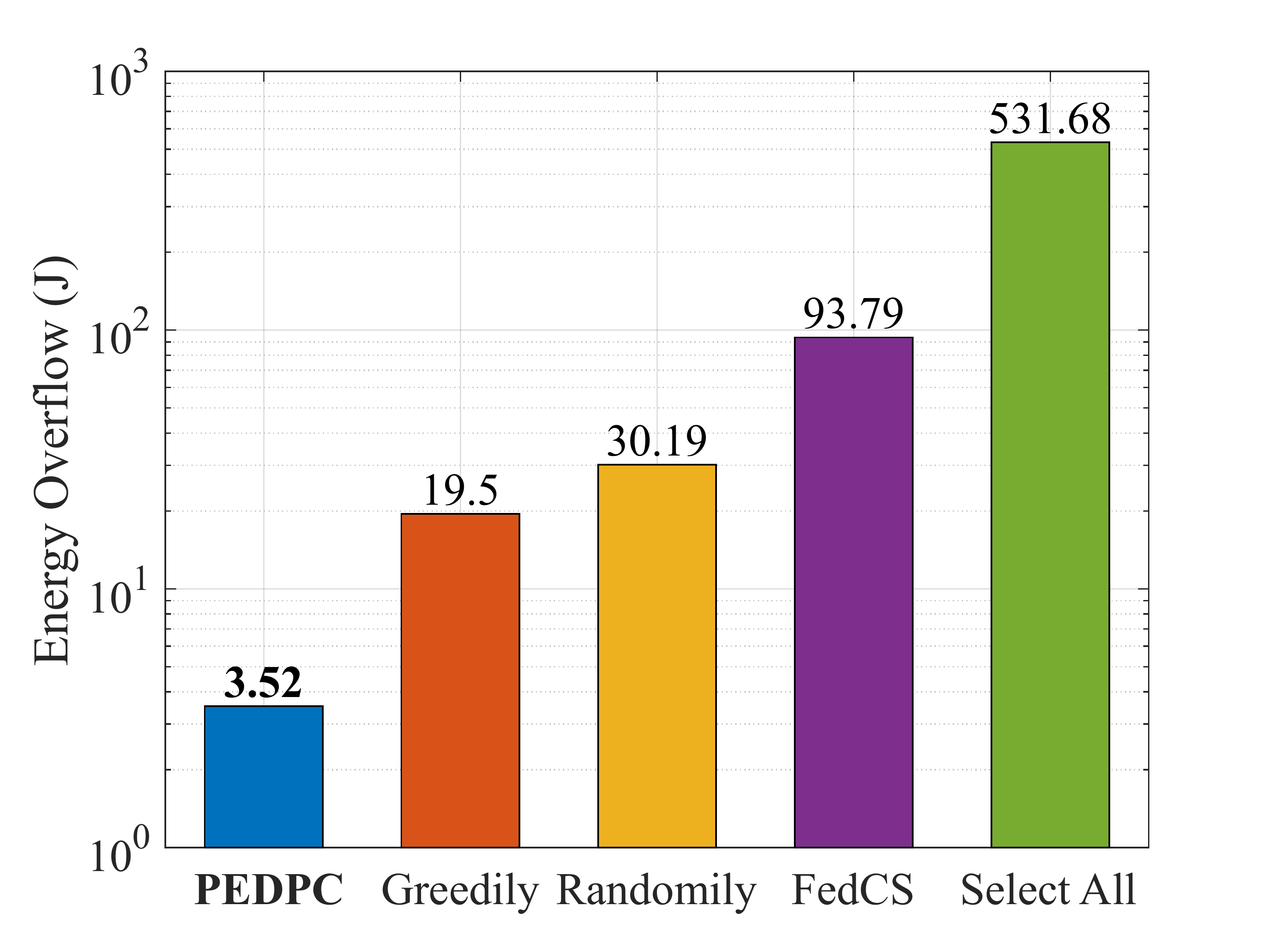}\label{fig: 4}}
\quad
\subfigure[Time Latency]{\includegraphics[width=0.3\textwidth]{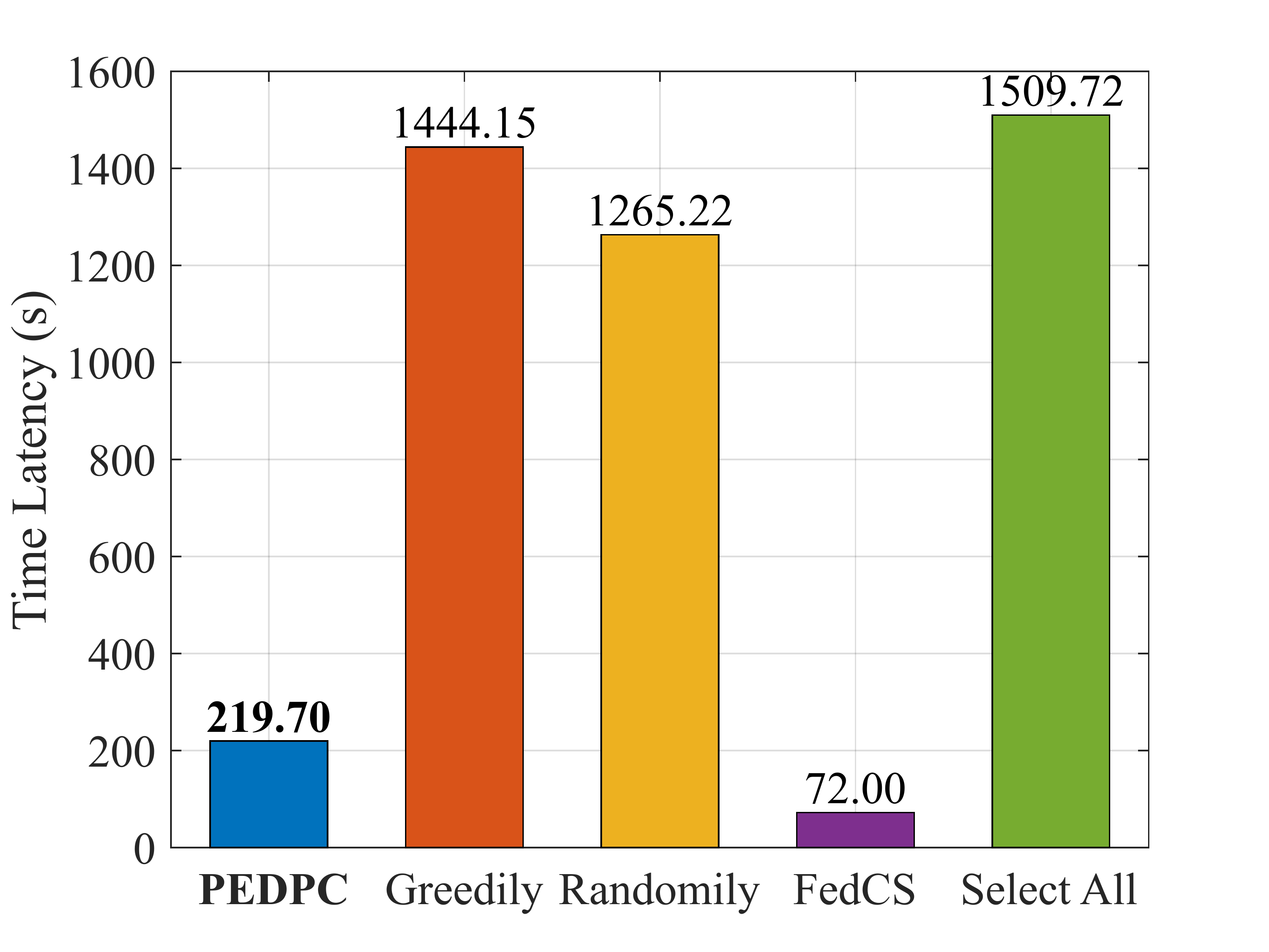}\label{fig: 5}}
\quad
\subfigure[Test accuracy]{\includegraphics[width=0.3\textwidth]{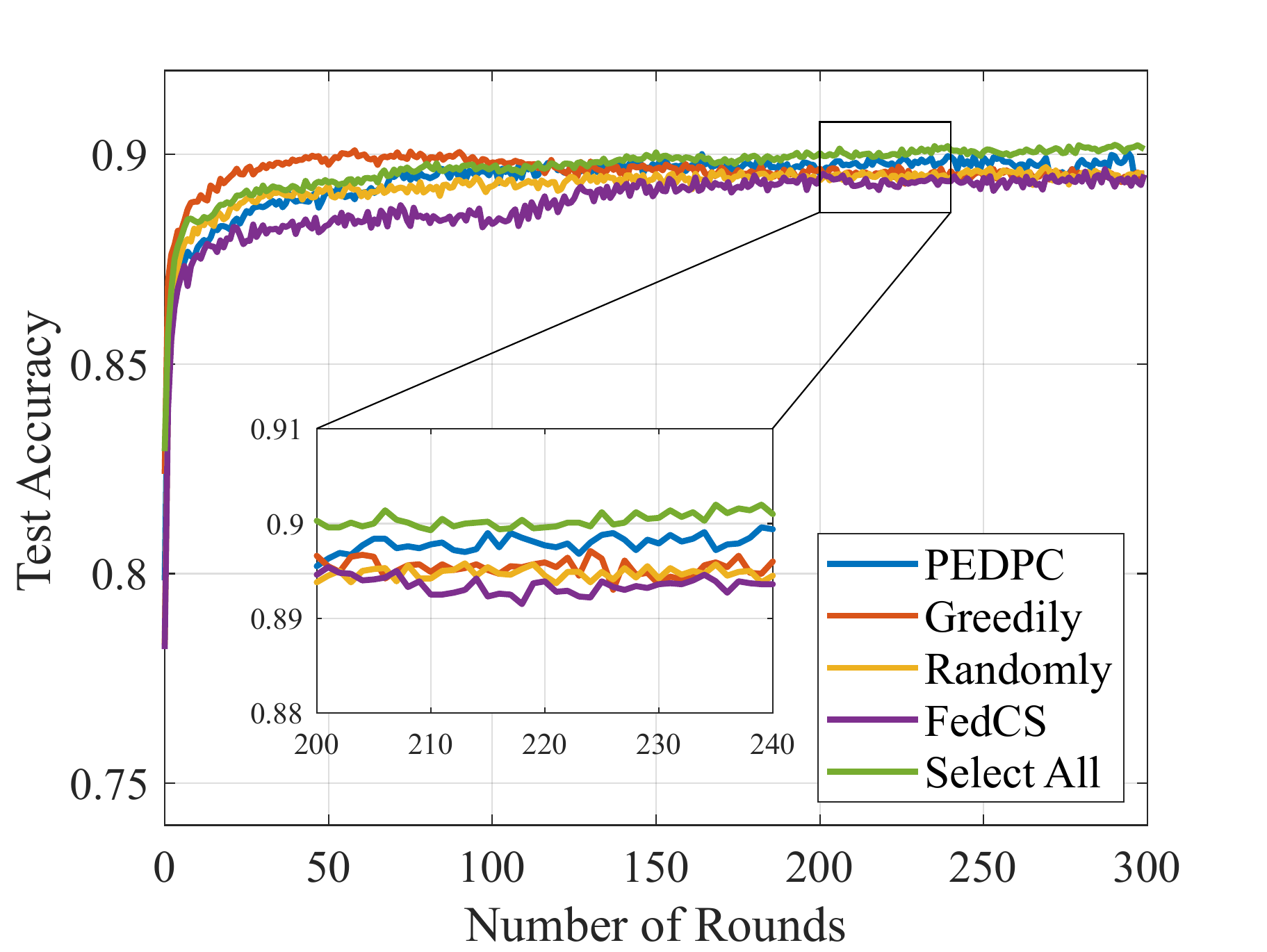}\label{fig: 6}}
\caption{The energy overflow, time latency and test accuracy of PEDPC and benchmarks in IID case}
\label{fig2}
\end{figure*}%

\begin{figure*}[htb]
\centering
\subfigure[Energy overflow]{\includegraphics[width=0.3\textwidth]{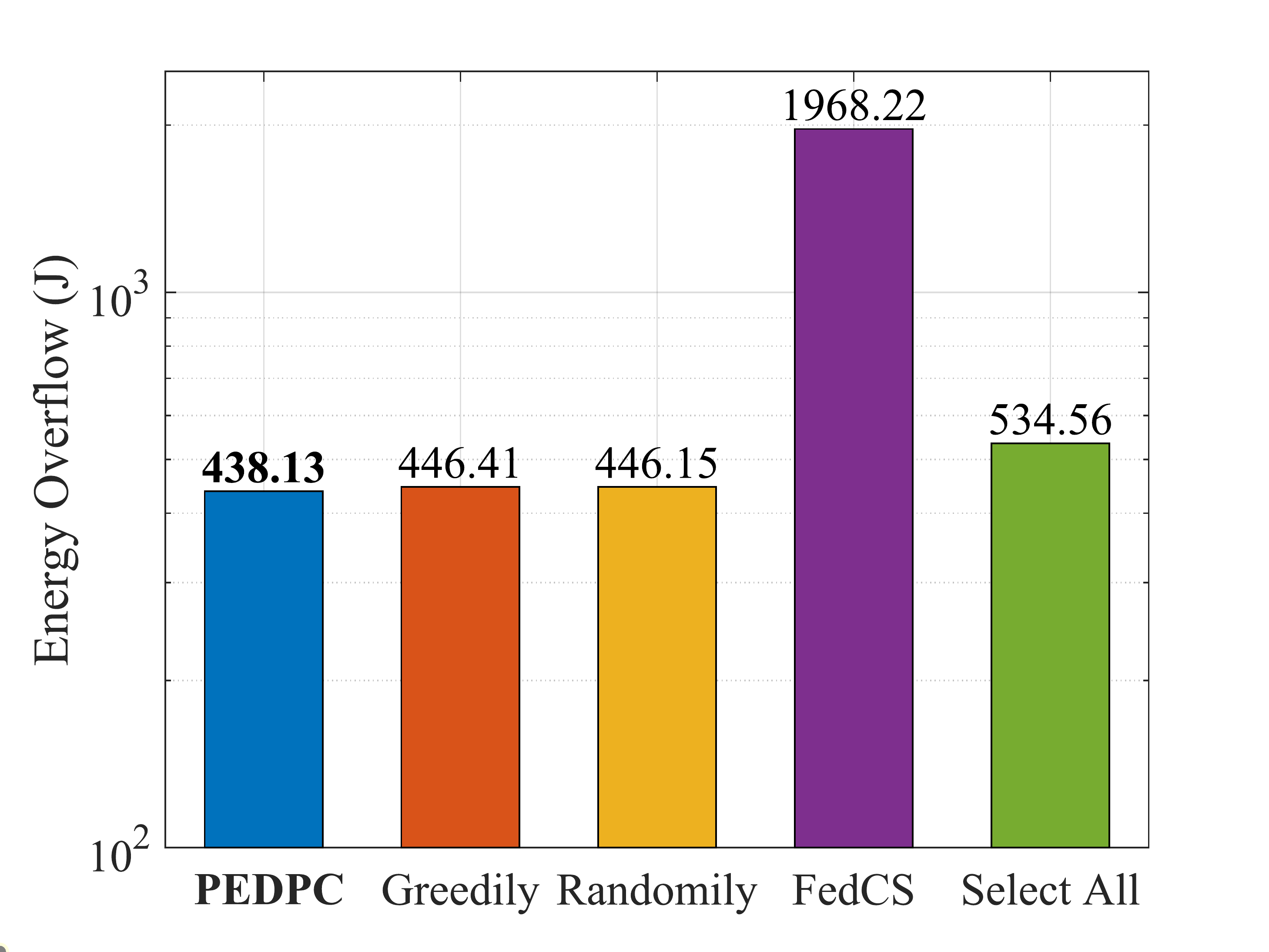}\label{fig: 7}}
\quad
\subfigure[Time Latency]{\includegraphics[width=0.3\textwidth]{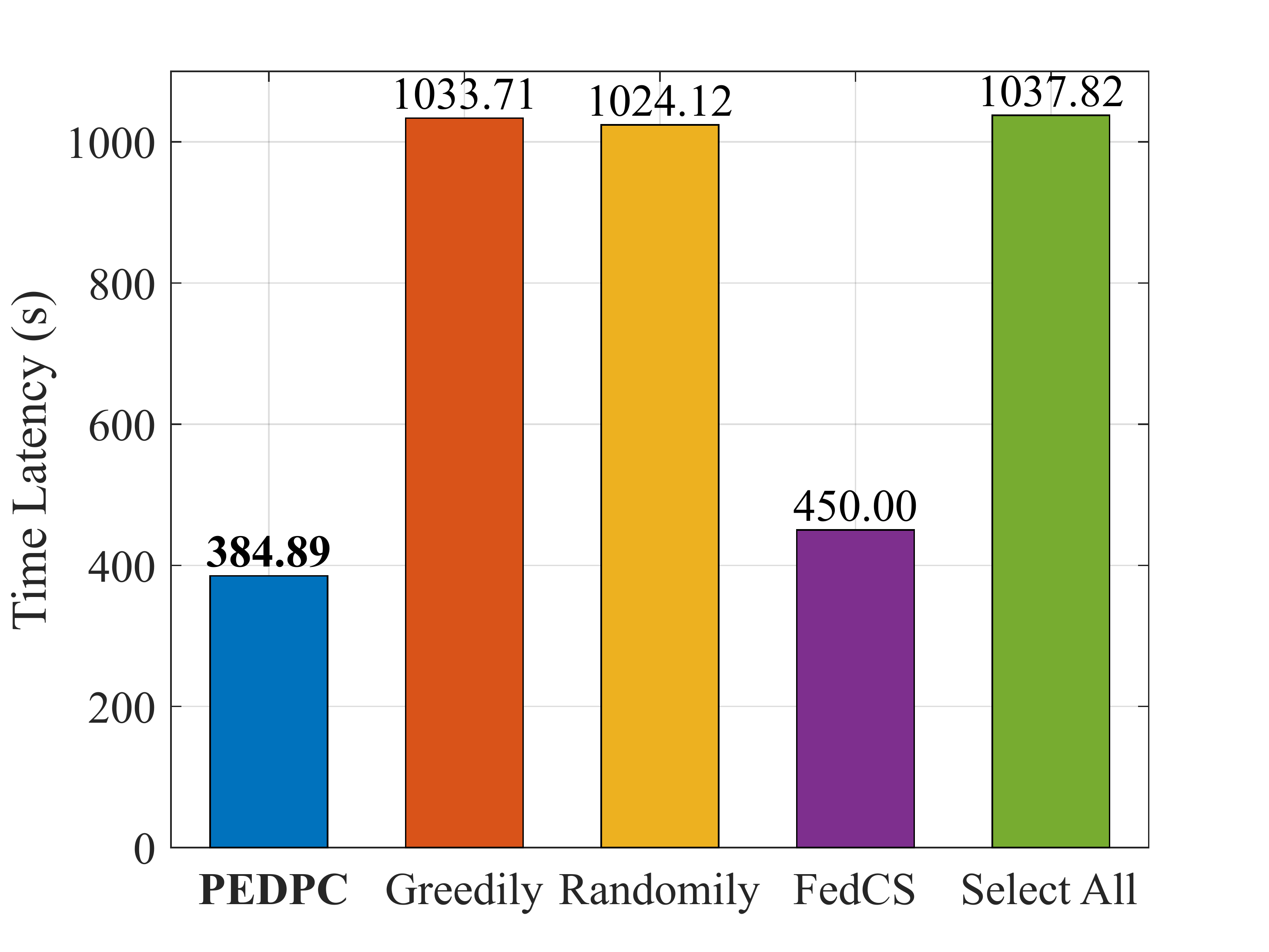}\label{fig: 8}}
\quad
\subfigure[Test accuracy]{\includegraphics[width=0.3\textwidth]{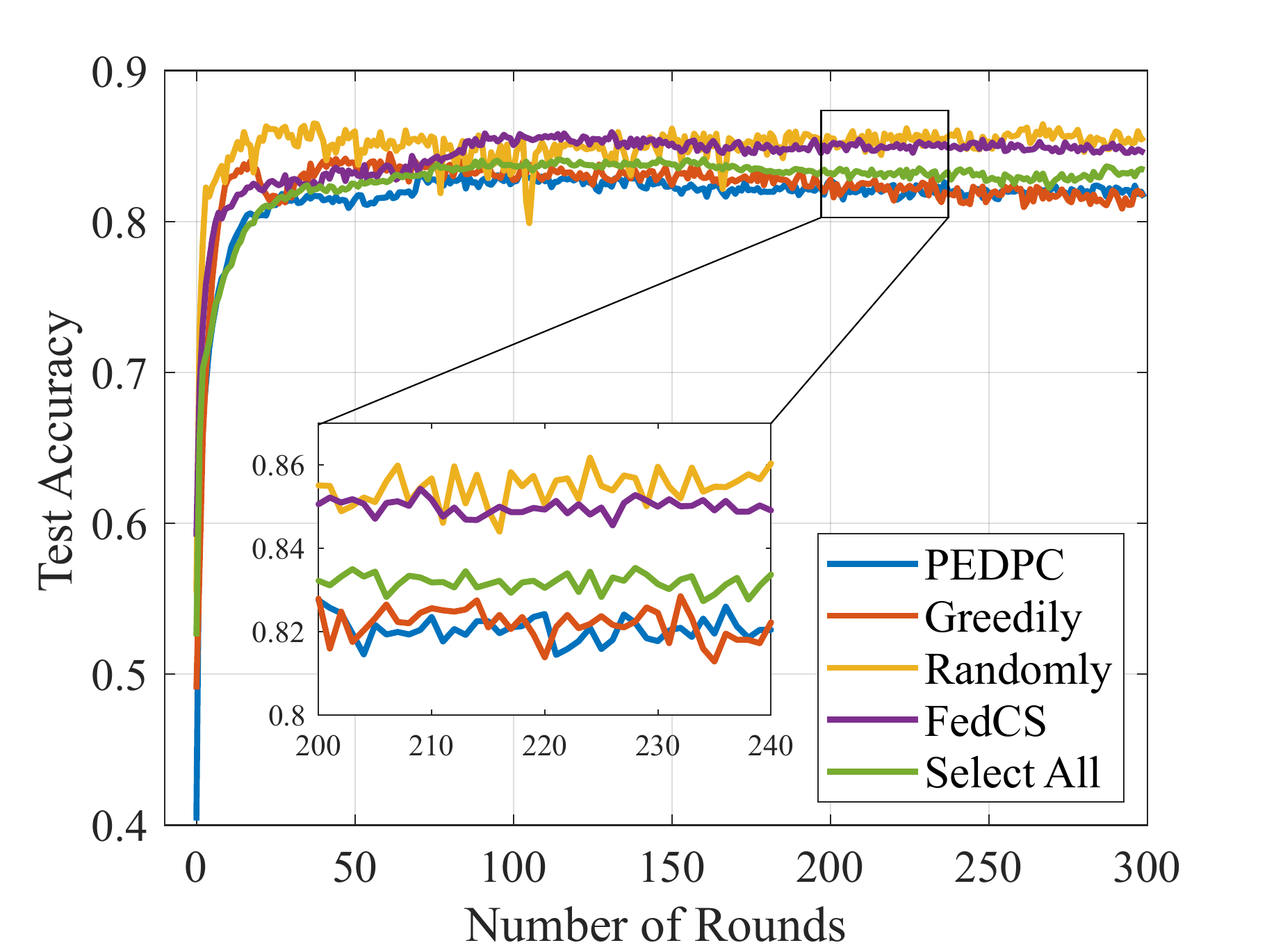}\label{fig: 9}}
\caption{The energy overflow, time latency and test accuracy of PEDPC and benchmarks in NON-IID case}
\label{fig3}
\end{figure*}

\clearpage
\section*{Appendix}
\subsection{Proof of Lemma 1}
Plugging (18) and (19) in energy drift function yields:
\begin{equation}
\begin{aligned}
    &Y(Z(r+1))-Y(Z(r))\\
    &=\frac{1}{2}\sum\limits_{k=1}^{K}{\left( {{\left( {{Z}_{k}}(r)+{{x}_{k}}(r){{E}_{k}}(r)-{{H}_{k}}/R \right)}^{2}}-{{Z}_{k}}{{(r)}^{2}} \right)} \\
    &=\frac{1}{2}\sum\limits_{k=1}^{K}{{{\left( {{x}_{k}}(r){{E}_{k}}(r)-{{H}_{k}}/R \right)}^{2}}}\\
    &\quad+\sum\limits_{k=1}^{K}{{{Z}_{k}}(r)}({{x}_{k}}(r){{E}_{k}}(r)-{{H}_{k}}/R)\\
    &\le D+\sum\limits_{k=1}^{K}{{{Z}_{k}}(r)}({{x}_{k}}(r){{E}_{k}}(r)-{{H}_{k}}/R)
\end{aligned}
\tag{33}
\end{equation}
where $D=\frac{1}{2}\sum\limits_{k=1}^{K}{\max [{{(y_{k}^{\min })}^{2}},{{(y_{k}^{\max })}^{2}}]}$.
\subsection{Proof of Theorem 1}
Assume $g(x)=\ln ({{e}^{{{x}_{1}}}}+{{e}^{{{x}_{2}}}}+...+{{e}^{{{x}_{m}}}})$, $f(b)=\sum\limits_{k\in {{S}^{*}}}^{{}}{\frac{G_{k}^{'}}{{{b}_{k}}}}$, $h\left( b \right)$: $R_{+}^{m}\to R_{+}^{m}$, $h\left( b \right)={{[{{C}_{1}}+\frac{S_{1}^{'}}{{{b}_{1}}},{{C}_{2}}+\frac{S_{2}^{'}}{{{b}_{2}}},...,{{C}_{m}}+\frac{S_{m}^{'}}{{{b}_{3}}}]}^{T}}$, $m=|{{S}^{*}}|$, then we rewrite P6 as follows:
\begin{equation}
\begin{aligned}
    & \underset{b(r)}{\mathop{\min }}\,Vg\left( h\left( b \right) \right)+f(b) \\
    & \text{s.t.}\text{ }\text{Constraints}(13),(14) \\ 
\end{aligned}
\tag{34}
\end{equation}

We denote $\boldsymbol{D}$ as the feasible region of the above problem, and obviously $\boldsymbol{D}$ is a convex set. There have two steps to proof P6 is a convex function:
1)	proof $g(x)$ is a convex function:
\begin{equation}
\begin{aligned}
    &\frac{\partial g(x)}{\partial {{x}_{i}}}=\frac{{{e}^{{{x}_{i}}}}}{({{e}^{{{x}_{1}}}}+{{e}^{{{x}_{2}}}}+...+{{e}^{{{x}_{m}}}})},\forall i\in 1,2,...m \\
    &\frac{\partial g(x)}{\partial {{x}_{i}}\partial {{x}_{j}}}=\left\{ \begin{matrix}
   \frac{-{{e}^{{{x}_{i}}}}{{e}^{{{x}_{j}}}}}{{{({{e}^{{{x}_{1}}}}+{{e}^{{{x}_{2}}}}+...+{{e}^{{{x}_{m}}}})}^{2}}} & i\ne j  \\
   \frac{{{e}^{{{x}_{i}}}}({{e}^{{{x}_{1}}}}+{{e}^{{{x}_{2}}}}+...+{{e}^{{{x}_{m}}}})-{{e}^{2{{x}_{i}}}}}{{{({{e}^{{{x}_{1}}}}+{{e}^{{{x}_{2}}}}+...+{{e}^{{{x}_{m}}}})}^{2}}} & i=j  \\
\end{matrix} \right.
\end{aligned}
\tag{35}
\end{equation}

Let $z={{({{e}^{{{x}_{1}}}},{{e}^{{{x}_{2}}}},...,{{e}^{{{x}_{m}}}})}^{T}}$, then ${{\nabla }^{2}}g(x)=H=\frac{1}{{{({{1}^{T}}z)}^{2}}}\left( ({{1}^{T}}z)diag\left\{ z \right\}-z{{z}^{T}} \right)$, for $\forall v\in {{R}^{m}}$, we have:
\begin{equation}
\begin{aligned}
    & {{v}^{T}}\left[ ({{1}^{T}}z)diag\left\{ z \right\}-z{{z}^{T}} \right]v \\ 
    & =({{1}^{T}}z){{v}^{T}}diag\left\{ z \right\}v-{{v}^{T}}z{{z}^{T}}v \\ 
    & =\left( \sum\limits_{i=1}^{m}{{{z}_{i}}} \right)\left( \sum\limits_{i=1}^{m}{{{v}_{i}}^{2}{{z}_{i}}} \right)-{{\left( \sum\limits_{i=1}^{m}{{{v}_{i}}{{z}_{i}}} \right)}^{2}} \\ 
    & ={{b}^{T}}b({{a}^{T}}a)-{{({{a}^{T}}b)}^{2}} \\ 
    & \ge 0
\end{aligned}
\tag{36}
\end{equation}
where ${{a}_{i}}={{v}_{i}}\sqrt{{{z}_{i}}},{{b}_{i}}=\sqrt{{{z}_{i}}}$, $a=({{a}_{1}},{{a}_{2}},...,{{a}_{m}})$, $b=({{b}_{1}},{{b}_{2}},...,{{b}_{m}})$ and the last inequality is due to the Cauchy–Schwarz inequality. Thus, $H\succcurlyeq 0$, and $g(x)$ is a convex function.

2)	Proof $h\left( b \right):R_{+}^{m}\to R_{+}^{m}$, $h\left( b \right)={{[{{h}_{1}},{{h}_{2}},...,{{h}_{m}}]}^{T}}={{[{{C}_{1}}+\frac{S_{1}^{'}}{{{b}_{1}}},{{C}_{2}}+\frac{S_{2}^{'}}{{{b}_{2}}},...,{{C}_{m}}+\frac{S_{m}^{'}}{{{b}_{3}}}]}^{T}}$ is a convex function.

For $\forall x=\{{{x}_{1}},{{x}_{2}},...,{{x}_{m}}\}\in D$, $y=\{{{y}_{1}},{{y}_{2}},...,{{y}_{m}}\}\in D$, $\theta \in [0,1]$, $\forall i=1,2,...,m$, we have:

\begin{equation}
\begin{aligned}
    & \theta \left( {{C}_{i}}+\frac{S_{i}^{'}}{{{x}_{i}}} \right)+\left( 1-\theta  \right)\left( {{C}_{i}}+\frac{S_{i}^{'}}{{{y}_{i}}} \right) \\ 
    & ={{C}_{i}}+{{S}_{i}}\left( \frac{\theta }{{{x}_{i}}}+\frac{1-\theta }{{{y}_{i}}} \right) \\ 
    & \ge {{C}_{i}}+{{S}_{i}}\left( \frac{1}{\theta {{x}_{i}}+(1-\theta ){{y}_{i}}} \right)
\end{aligned}
\tag{37}
\end{equation}

Thus ${{h}_{i}},\forall i$ is a convex function, Because of $V>0$ and $\frac{\partial g(x)}{\partial {{x}_{i}}}\ge 0$, $\forall i\in 1,2,...m$, $Vg(x)$ is convex and nondecreasing in each argument. According to the \textbf{Vector composition} $Vg\left( h\left( b \right) \right)$ is a convex function. Besides, ${{\nabla }^{2}}f(x)=diag\left\{ \frac{2G_{1}^{'}}{b_{1}^{3}},\frac{2G_{2}^{'}}{b_{2}^{3}},...,\frac{2G_{m}^{'}}{b_{m}^{3}} \right\}\succcurlyeq 0$,  so P8 is a convex problem.

\subsection{Proof of Lemma 2}
We rewrite (18) as follows:
\begin{equation}
\begin{aligned}
    & {{Z}_{k}}(r+1)\\
    &=\max [{{Z}_{k}}(r)+{{x}_{k}}(r){{E}_{k}}(r)-{{H}_{k}}/R,\text{ }0] \\ 
    &=\max [{{Z}_{k}}(r)-(-{{x}_{k}}(r){{E}_{k}}(r)+{{H}_{k}}/R),\text{ }0] \\ 
    &={{Z}_{k}}(r)-\min [-{{x}_{k}}(r){{E}_{k}}(r)+{{H}_{k}}/R,\text{ }{{Z}_{k}}(r)]  
\end{aligned}
\tag{38}
\end{equation}
for any $r\ge 0$ we have :
\begin{equation}
\begin{aligned}
   &{{Z}_{k}}(r+1)-{{Z}_{k}}(r)\\
   &=-\min [-{{x}_{k}}(r){{E}_{k}}(r)+{{H}_{k}}/(LF),\text{ }{{Z}_{k}}(r)]
\end{aligned}
\tag{39}
\end{equation}

Summing (39) over $r\in \{0,1,...R-1\}$, we have:
\begin{equation}
\begin{aligned}
    & {{Z}_{k}}(R)-{{Z}_{k}}(0)\\
    &=-\sum\limits_{r=0}^{R-1}{\min [-{{x}_{k}}(r){{E}_{k}}(r)+{{H}_{k}}/R,\text{ }{{Z}_{k}}(r)]} \\ &\ge \sum\limits_{r=0}^{R-1}{\left( {{x}_{k}}(r){{E}_{k}}(r)-{{H}_{k}}/R \right)} \\ 
    &=\sum\limits_{r=0}^{R-1}{{{x}_{k}}(r){{E}_{k}}(r)-{{H}_{k}}} 
\end{aligned}
\tag{40}
\end{equation}
\subsection{Proof of Theorem 2}
Let ${{E}_{k}}(r)=\widehat{{{E}_{k}}}({{b}_{k}}(r))$, and assume ${{x}^{*}}(r)$, ${{b}^{*}}(r)$ are the optimal solution of P2 by L-round lookahead algorithm. According to Lemma 4.11 of [11], we have:
\begin{equation}
\begin{aligned}
    & Y(Z(fL+L))-Y(Z(fL))+V\sum\limits_{r=fL}^{fL+L-1}{{{y}_{0}}(x(r),b(r))} \\ 
    & \le D{{L}^{2}}+V\sum\limits_{r=fL}^{fL+L-1}{{{y}_{0}}(x(r),b(r))}\\
    &\quad +\sum\limits_{k=1}^{K}{{{Z}_{k}}}(fL)\sum\limits_{r=fL}^{(f+1)L-1}{\left(x_{k}^{*}(r)\widehat{{{E}_{k}}}(b_{k}^{*}(r))-{{H}_{k}}/R \right)} \\ 
    & \le D{{L}^{2}}+VLc_{f}^{*} 
\end{aligned}
\tag{41}
\end{equation}
where the last inequality is because ${{x}^{*}}(r)$, ${{b}^{*}}(r)$ satisfy the constraint (17) of P2 and ${{Z}_{k}}(fL)\ge 0$, $\forall k,f\in \{0,1,cdots,F-1\}$.

Summing the above over $f\in \{0,\cdots,F-1\}$(for any integer $F>0$) yields:
\begin{equation}
\begin{aligned}
    &Y(Z(FL))-Y(Z(0))+V\sum\limits_{r=0}^{FL-1}{{{y}_{0}}(x(r),b(r))}\\
    &\le D{{L}^{2}}F+VL\sum\limits_{f=0}^{F-1}{c_{f}^{*}}
\end{aligned}
\tag{42}
\end{equation}

Dividing by $VLF$, using the fact that $Y(Z(FL))\ge 0$, plugging that $R=FL$, and rearranging terms yields:
\begin{equation}
\begin{aligned}
    \frac{1}{R}\sum\limits_{r=0}^{FL-1}{{{y}_{0}}(x(r),b(r))}\le \frac{1}{F}\sum\limits_{f=0}^{F-1}{c_{f}^{*}}+\frac{DL}{V}+\frac{Y(Z(0))}{VR}
\end{aligned}
\tag{43}
\end{equation}

When $Y(Z(0))=0$, we have that the objective of P1 is within $O(1/V)$ of the optimal value.
\subsection{Proof of Theorem 3}
Rearranging terms of (42) we have:
\begin{equation}
\begin{aligned}
    &Y(Z(FL))-Y(Z(0))\\
    &\le D{{L}^{2}}F+V\left( L\sum\limits_{f=0}^{F-1}{c_{f}^{*}}-\sum\limits_{r=0}^{FL-1}{{{y}_{0}}(x(r),b(r))} \right)
\end{aligned}
\tag{44}
\end{equation}
Plugging $V\left( L\sum\limits_{f=0}^{F-1}{c_{f}^{*}}-\sum\limits_{r=0}^{FL-1}{{{y}_{0}}(x(r),b(r))} \right)\le VL(c_{f}^{*}-y_{0}^{\min })$ in to (44) yields:
\begin{equation}
\begin{aligned}
    Y(Z(FL))\le D{{L}^{2}}F+VL\sum\limits_{f=0}^{F-1}{(c_{f}^{*}}-y_{0}^{\min })+Y(Z(0))
\end{aligned}
\tag{45}
\end{equation}
Due to $Y(Z(FL))=\frac{1}{2}\sum\limits_{k=1}^{K}{{{Z}_{k}}{{(FL)}^{2}}}$ we have:
\begin{equation}
\begin{aligned}
    & {{Z}_{k}}{{(FL)}^{2}} \\ 
    & \le \sum\limits_{k=1}^{K}{{{Z}_{k}}{{(FL)}^{2}}} \\ 
    & =2Y(Z(FL)) \\ 
    & \le 2D{{L}^{2}}F+2VL\sum\limits_{f=0}^{F-1}{(c_{f}^{*}}-y_{0}^{\min })+2Y(Z(0)),\forall k\in \mathcal{K}  
\end{aligned}
\tag{46}
\end{equation}
Dividing by $FL$ yields:
\begin{equation}
\begin{aligned}
    & \frac{{{Z}_{k}}(FL)}{FL} \\ 
    & \le \sqrt{\frac{2D}{F}+\frac{2V\sum\limits_{f=0}^{F-1}{(c_{f}^{*}}-y_{0}^{\min })}{{{F}^{2}}L}+\frac{2Y(Z(0))}{{{F}^{2}}{{L}^{2}}}} 
\end{aligned}
\tag{47}
\end{equation}
Taking limits of both sides of the above and using the non-negativity of ${{Z}_{k}}(r)$ yields:
\begin{equation}
\begin{aligned}
   \underset{r\to \infty }{\mathop{\lim }}\,\frac{{{Z}_{k}}(r)}{r}=0
\end{aligned}
\tag{48}
\end{equation}
According to Lemma 1, we have:
\begin{equation}
\begin{aligned}
   {{Z}_{k}}(FL)\ge \sum\limits_{r=0}^{R-1}{{{x}_{k}}(r){{E}_{k}}(r)}-{{H}_{k}}
\end{aligned}
\tag{49}
\end{equation}
Plugging (47) into (49) and assuming $Y(Z(0))=0$ yields:
\begin{equation}
\begin{aligned}
   \sum\limits_{r=0}^{R-1}{{{x}_{k}}(r){{E}_{k}}(r)}\le {{H}_{k}}+\sqrt{2DRL+2VL\sum\limits_{f=0}^{F-1}{(c_{f}^{*}}-y_{0}^{\min })}
\end{aligned}
\tag{50}
\end{equation}
Thus the energy consumption of each client is upper bounded by ${{H}_{k}}+\sqrt{2DRL+2VL\sum\limits_{f=0}^{F-1}{(c_{f}^{*}}-y_{0}^{\min })}$.

\vspace{12pt}

\end{document}